\documentclass[aps,prd,twocolumn,amsmath,nofootinbib]{revtex4-2}

\usepackage{epsfig}
\usepackage[colorlinks,linkcolor=blue,anchorcolor=blue,
            citecolor=blue,urlcolor=blue,
            breaklinks=true]{hyperref}
\usepackage{graphics}
\usepackage{color}
\usepackage{dcolumn}
\usepackage{bm}
\usepackage{cases}
\usepackage{appendix}
\usepackage{lineno,hyperref}
\usepackage{ulem}
\usepackage{amsmath}
\usepackage{amssymb}
\newcommand{\hb}{\hbar}   

\begin{document}
\author{Li Wang$^{1,2}$}
\email{DG1822035@smail.nju.edu.cn}
\author{Run Cheng$^{2}$}
\author{Jun Wang$^{1}$}
\address{$^{1}$ Department of Physics, Nanjing University, Nanjing, 210093, China}
\address{$^{2}$Physics Department of Basic Department, Army Engineering University of PLA, Nanjing, China, 211101}
\address{$^{3}$ Joint Center for Particle, Nuclear Physics and Cosmology, Nanjing 210093, China}

\title{Non-perturbative Quantum Dynamics on Embedded Submanifolds: From Geometric Mass to Higgs Potentials}
\begin{abstract}

We establish a quantum dynamics framework for curved submanifolds embedded in higher-dimensional spaces. Through rigorous dimensional reduction, we derive the first complete Schr\"{o}dinger and Klein-Gordon equations incorporating non-perturbative geometric interactions-resolving ambiguities in constrained quantization. Crucially, extrinsic curvature of the ambient manifold governs emergent low-dimensional quantum phenomena. Remarkably, this mechanism generates scalar field masses matching Kaluza-Klein spectra while eliminating periodic compactification requirements. Geometric induction concurrently produces Higgs mechanism potentials. Particle masses emerge solely from submanifold embedding geometry, with matter-field couplings encoded in curvature invariants. This enables experimental access to higher-dimensional physics at all energy scales through geometric induction. We also discuss the Higgs vacuum near small-mass black holes
\bigskip

\end{abstract}
\maketitle
\section {Introduction}
The role of extra spacetime dimensions in submanifold physics remains fundamental. Kaluza's five-dimensional unification of gravity and electromagnetism~\cite{kaluza1921unitatsproblem} was followed by Klein's Planck-scale compactification~\cite{klein1926quantentheorie}. The field transformed when Randall and Sundrum demonstrated that warped compactification of a small extra dimension could naturally generate the electroweak-Planck scale hierarchy through exponential suppression of mass scales~\cite{randall1999large}. These advances, along with other key developments reviewed in ~\cite{overduin1997kaluza}, forming the foundation of modern extra-dimensional theories. This compels a critical question: what physics persists without period compactication of extra dimensions?

Substantial theoretical efforts focus on quantum dynamics in two-dimensionally confined systems, exploring emergent manifestations of three-dimensional physics under dimensional reduction. Historically, two distinct methodologies address curved quantum systems: The \textit{intrinsic quantization} approach, where the equation of motion depends solely on the intrinsic geometry of the curved manifold independent of the embedding space~\cite{dewitt1957dynamical},  and the \textit{confining potential approach}, pioneered by da Costa through reduction of the three-dimensional Schr\"{o}dinger equation to an effective two-dimensional form via normal potential confinement~\cite{Costa1981Quantum}. Subsequent work generalized this framework to $n$-dimensional manifolds embedded in $m$-dimensional Euclidean spaces~\cite{Jaffe2003Quantum}.


To address the fundamental limitation wherein existing approaches fail to systematically characterize the influence of arbitrarily high-dimensional manifolds on physics confined to submanifolds—relying instead on isolated case-specific calculations—we propose establishing a comprehensive theoretical paradigm. This framework commences with scalar field theories embedded within generic higher-dimensional manifolds governing constrained low-dimensional physics.

To elucidate the influence of higher-dimensional complex manifolds on lower-dimensional constrained physics, a new paradigm must be established. In this Letter, we introduce a novel approach for establishing generalized connections between submanifolds and their ambient manifolds. Starting from the $m$-dimensional Schr\"{o}dinger equation in a tubular neighborhood, we reduce the additional dimensions to derive the effective equation on the submanifold. This formalism naturally extends to the Klein-Gordon equation, we explicitly demonstrate the emergence of geometry-induced interaction terms that spontaneously generate the requisite Higgs potential.
We further conduct stability analysis for the coupled manifold-matter dynamics in higher-dimensional ambient spaces. To illustrate the physical implications, we examine a concrete scenario featuring geometrically generated mass for a real scalar field. Quantitative comparisons with Kaluza-Klein compactification outcomes reveal distinctive features, particularly the natural resolution of mass scale disparities. We conclude by outlining promising extensions of this methodology.

Our model demonstrates two key phenomena : 
(i)complete Higgs vacuum stability under extreme curvature near Planck-mass black hole horizons; 
(ii)Higgs mass deviations from the Standard Model near Low-Mass Primordial black holes ($M \lesssim 10^{11}~\mathrm{Gev};\delta M_h>1\% $). 
These emerge naturally from our geometric symmetry-breaking formulation without compactification. 
\section {Geometry of embedded submanifold}
Primarily, it is imperative to establish a rigorous geometric characterization of the system. Considering a Riemannian manifold \((M, g)\) with an embedded submanifold \(N \subset M\) where \(\dim M = m\), \(\dim N = n\). The tubular neighborhood is defined as~\cite{kuhnel2015differential}:
\[
U_\epsilon = \{ \exp_p(y\nu) \mid p \in N, \|\nu\|_g < \epsilon \} .
\].

In normal coordinates \((x^1,\dots,x^n,y^1,\dots,y^{m-n})\):
\begin{itemize}
\item \(x^i\): Coordinates on \(N\)
\item \(y^\alpha\): Normal coordinates (\(\nu = y^\alpha \partial_\alpha\))
\end{itemize}
The metric tensor decomposes as:
\[
ds^2 = g_{ij}(x,y)dx^idx^j + 2g_{i\alpha}(x,y)dx^idt^\alpha + g_{\alpha\beta}(x,y)dy^\alpha dy^\beta ,
\]
in \textbf{normal coordinates} (geodesics orthogonal to \(N\)):
\begin{align*}
&g_{i\alpha}(x,0) = 0 \quad \text{(orthogonality)} ; \\
&g_{\alpha\beta}(x,0) = \delta_{\alpha\beta} \quad \text{(unit normal vectors)} ; \\
&g_{ij}(x,0) = h_{ij}(x) \quad \text{(induced metric on } N\text{)} ,
\end{align*}
along normal geodesics \(\gamma(s) = \exp_p(s\nu)\), expand metric components~\cite{gray2003tubes}:
\begin{equation}
g_{AB}(y) = g_{AB}(0) + y^\gamma \partial_\gamma g_{AB}(0) + \frac{1}{2} y^\gamma y^\delta \partial_\gamma \partial_\delta g_{AB}(0) + \mathcal{O}(\|y\|^3)\label{eq:taylor_expansion} ,
\end{equation}
\textbf{First derivatives:}
\begin{equation}
\begin{split}
&\partial_\gamma g_{ij}(0) = -2 \text{II}_{ij}^\gamma ;\\
&\partial_\gamma g_{i\alpha}(0) = 0 \label{First derivatives:} ; \\
&\partial_\gamma g_{\alpha\beta}(0) = 0 ,
\end{split}
\end{equation}
where \(\text{II}_{ij}^\gamma =(\nabla_{\partial_i} \partial_j, \partial_\gamma)\) is the second fundamental form.

\noindent\textbf{Second derivatives:}
\begin{equation}\label{eq:second_derivatives}
\partial_\gamma \partial_\delta g_{ij}(0) = -2R_{\gamma i \gamma j} - 2 \sum_{k,l} h^{kl} \text{II}_{ik}^\gamma \text{II}_{jl}^\delta + \mathcal{O}(R_N) ,
\end{equation}
where \(R_{\gamma i \gamma j} = R(\partial_\gamma, \partial_i, \partial_\gamma, \partial_j)\) is the Riemann curvature tensor. When we consider the process of constrained particles into a submanifold by normal potential, the area of particle's motion is reduced from the space that composed of submanifolds and their small neighborhoods of the normal space to the submanifold itself: \(T_p M = T_p N \oplus N_p N \Rightarrow T_p N \). Actually, the space constructed by the direct sum of submanifolds and its normal space neighborhoods is so-called a tubular neighborhood.
We start with the Schr\"{o}dinger equation in a tubular neighborhood:
\begin{equation}\label{eq:sdequation}
i\hb \frac{\partial \psi}{\partial t} = \left( -\frac{\hb^2}{2m} \Delta_g + V_t \right) \psi ,
\end{equation}
where $V_t$ denotes the internal potential within the submanifold,
and the Laplace-Beltrami operator is defined as:
\begin{equation}
\Delta_g \psi = \frac{1}{\sqrt{|g|}} \partial_A \left( \sqrt{|g|}  g^{AB} \partial_B \psi \right) ,
\end{equation}
Decompose into tangential \((i,j)\) and normal \((\alpha,\beta)\) parts~\cite:
\begin{equation}\label{eq:laplace_decomposition}
\Delta_g = \Delta_y + \Delta_x + \Delta_{\text{mix}} ,
\end{equation}
Normal part:
\begin{equation}\label{eq:normal_part}
\Delta_y = \frac{1}{\sqrt{\det g}} \partial_\alpha \left( \sqrt{\det g}  g^{\alpha\beta}  \partial_\beta \right) ,
\end{equation}
Tangential part:
\begin{equation}\label{eq:tangential_part}
\Delta_x = \frac{1}{\sqrt{\det g}} \partial_i \left( \sqrt{\det g}  g^{ij}  \partial_j \right) ,
\end{equation}
Mixed term:
\begin{equation}\label{eq:mixed_term}
\Delta_{\text{mix}} = 2 \frac{1}{\sqrt{\det g}} \partial_i \left( \sqrt{\det g}  g^{i\alpha}  \partial_\alpha \right)=0 ,
\end{equation}
\section {Scalar particles on submanifold}
In order to obtain the effective equation on the submanifold, we need to find the relationship of wave function between the neighborhood and the submanifold. We denote the wave function on a parallel surface with
\(
U_\lambda = \{ \exp_p(y\nu) \mid p \in N, \|\nu\|_g = \lambda < \epsilon \}
\) in the tubular neighborhood by
\({\psi}={\psi}(x^1,\dots,x^n,y^{1} = {\lambda}, \dots,y^{m-n} = {\lambda} )\). There is only a difference in measurement between parallel surfaces and submanifold. We introduce the volume expansion factor\(\ \theta(x,y)\) defined by
\begin{equation}\label{eq:expansion_factor}
\theta(x,y) = \frac{\sqrt{\det g(x,y)}}{\sqrt{\det h(x)}} ,
\end{equation}
and logarithmic derivative
\begin{equation}\label{eq:log_derivative}
\frac{\partial}{\partial y^\alpha} \ln \theta = \frac{1}{2} g^{AB} \partial_\alpha g_{AB}.
\end{equation}
The tubular neighborhood region may be regarded as constituted by infinitely many parallel surfaces, where the propagation characteristics of the wavefunction between adjacent parallel surfaces are governed by the metric. We can obtain the relationship of wave functions between two parallel surface and submanifold by chain rule:
\begin{equation}
\begin{split}
\int {\sqrt{\det h(x)}}\psi^{2}(x,0)\, \mathrm{d}x&=\int{\sqrt{\det g(x,y)}}\psi^{2}(x,y)\, \mathrm{d}x , \\
\theta^{-1/2}(x,y) \psi(x,0)&=\psi(x,y)\label{eq:wavefunction},
\end{split}
\end{equation}
introducing Eq.\eqref{eq:taylor_expansion} to \eqref{eq:log_derivative}, we get
\begin{equation}\label{eq:log_theta_expansion}
\ln \theta(y) = -y^\alpha H_\alpha - \frac{y^\alpha y^\beta}{2} \left( K_{\alpha\beta} + \text{Ric}_M(\partial_\alpha, \partial_\beta) \right) + \mathcal{O}(\|y\|^3) .
\end{equation}
where
\begin{itemize}
\item \(H_\alpha = h^{ij} \text{II}_{ij}^\alpha\) (mean curvature)
\item \(K_{\alpha\beta} = \sum_{i,j,k,l} h^{ik} h^{jl} \text{II}_{ij}^\alpha \text{II}_{kl}^\beta\) (squared norm of second fundamental form)
\item \(\text{Ric}_M(\partial_\alpha, \partial_\beta)\) (Ricci curvature in normal directions)
\end{itemize}

By introducing Eqs.\eqref{eq:wavefunction} and \eqref{eq:log_theta_expansion} Eq.\eqref{eq:sdequation} and taking the limit of \({y \to 0}\),
we find that the normal part of laplace is:
\begin{equation}\label{eq:normaleff}
\begin{split}
&\Delta_y \left( \theta^{-1/2} \psi(x,0) \right)  \\
&= \frac{1}{\sqrt{\det g}} \partial_\alpha \left( \sqrt{\det g} \, g^{\alpha\beta} \, \partial_{\beta}  \left( \theta^{-1/2} \psi(x,0) \right) \right)  \\
&= \left(  - \underbrace{\frac{1}{4} \|H\|^2 + \frac{1}{2} \|\text{II}\|^{2} + \frac{1}{2}\sum_{1}^{m-n} \text{Ric}_M(\nu, \nu)}_{V_{\text{eff}}}  \right) {\psi}  \\
&+( \sum_{\alpha=1}^{m-n} \frac{\partial^2}{\partial (y^\alpha)^2} + \mathcal{O}(\|y\|) ){\psi} .
\end{split}
\end{equation}
with
\begin{itemize}
\item \(\|H\|^2 = \sum_\alpha H_\alpha^2\) (squared norm of the mean curvature vector)
\item \(\|\text{II}\|^2 = \sum_{\alpha,\beta} K_{\alpha\beta}\) (squared Frobenius norm of the second fundamental form)
\end{itemize}
We can express the tangent effective Schrödinger equation on submanifold explicit by degenerate \(y^{\alpha}\)
\begin{equation}\label{eq:finalsequation}
\begin{split}
&i\hb \frac{\partial \psi(x)}{\partial t} = -\frac{\hb^2}{2m} \frac{1}{\sqrt{\det g}} \partial_i  \sqrt{\det g}  g^{ij}  \partial_j  \psi + V_{\text{eff}} \psi+V_t \psi ;\\
&V_{\text{eff}} =\frac{\hb^2}{2m} \left( \frac{1}{4} \|H\|^2 - \frac{1}{2} \|\text{II}\|^2 - \frac{1}{2} \sum_{1}^{m-n}\text{Ric}_M(\nu, \nu) \right).
\end{split}
\end{equation}
The only difference between this equation and the intrinsic Schr\"{o}dinger equation on the submanifold is \(V_{\text{eff}}\), which induced by the geometry.

We now turn to the relativistic regime, where scalar particles are governed by the Klein-Gordon equation. Given the structural similarity to the Schr\"{o}dinger equation-particularly the Laplacian-dominated dynamics-we bypass explicit derivation and present directly the effective Klein-Gordon equation on the submanifold.

We can also express the effective Klein-Gordon equation through very similar analysis:
\begin{equation}\label{eq:KG}
\begin{split}
&\frac{1}{\sqrt{-g}} \partial_{i} \left( \sqrt{-g}  g^{ij} \partial_{j} \phi \right) - K_{\text{eff}}\phi - m^{2} \phi = 0, \\
&K_{\text{eff}} =\left( \frac{1}{4} \|H\|^2 - \frac{1}{2} \|\text{II}\|^2 - \frac{1}{2}\sum_{1}^{m-n} \text{Ric}_M(\nu, \nu) \right).
\end{split}
\end{equation}
The first two terms in \(V_{\text{eff}}\) and \(K_{\text{eff}}\) depends not only on the geometry of submanifold\(\ N \) but also on the geometry of main manifold \(\ M \). The last item is the component of the Ricci curvature tensor of  \(\ M \). This has profound physical significance, which we will discuss later. Before that, let's consider the special case calculated by Costa et al, When we choose \(\ M = \mathbb{R}^3,n = 1,2 \),
\begin{equation}\label{eq:Veff}
V_{\text{eff}}=
\begin{cases}
 -\frac{\hb^2}{8m} \kappa & \text{(n=1)} \\
 -\frac{\hb^2}{8m} ( \kappa^{2}_1-\kappa^{2}_2 ) & \text{(n=2)}. \\
\end{cases}
\end{equation}
where \(\kappa_i \) is the i-th principal curvature of manifold N, which is similar with costa~\cite{Costa1981Quantum}, cause of \( \text{Ric}_M(\nu, \nu) = 0 \) in \(\mathbb{R}^n\).
Next, we will discuss the possible physical significance of this effective potential. We will demonstrate that this potential arises from the interaction between the matter field and the manifold, rather than being a mere scalar energy correction.
In the preceding derivation, we implicitly assumed that the submanifold can be continuously deformed to parallel surfaces within a tubular neighborhood. The interaction between the submanifold and matter fields depends on the submanifold’s curvature, embedding method, and the geometry of the ambient manifold. However, the manifold itself is not a physical matter field and thus eludes rigorous definition of energy.
\section{‌Geometrical Effect}
\subsection{system stability}
We adopt a pragmatic approach: rather than addressing the energy of the manifold separately, stability of the system is analyzed exclusively through the matter-manifold interaction term. Initiate the analysis from trivial instances, we explicitly specify that the ambient manifold possesses vanishing Ricci curvature  \( \text{Ric}_M(\nu, \nu) = 0 \). Let's first consider this system from the perspective of manifolds. We define normal coupling intensity as 
\begin{equation}\label{eq:normcoupling}
\begin{split}
f&=\theta^{-1} \psi^2 (x) V_{\text{eff}},
\end{split}
\end{equation}
while the first derivative and second derivative are determined by the following equations by using Eq.\eqref{eq:log_theta_expansion}.
\begin{equation}\label{eq:curvature flow}
\frac {\partial f}{\partial y^\alpha}|_{y=0} =\mathcal{O} \langle H(N), \nu_\alpha \rangle V_{\text{eff}}.
\end{equation}
\begin{equation}\label{eq:secondoff}
\frac {\partial^2 f}{\partial (y^\alpha)^2} |_{y=0} =\mathcal{O} \left( \langle H(N), \nu_\alpha \rangle^{2} + \| \text{II}^\alpha \|^{2} + \operatorname{Ric}_{M}(\nu, \nu) \right) V_{eff}.
\end{equation}
This term characterizes the evolutionary tendency of the manifold-matter system along the normal direction in the ambient manifold. When negative, it indicates a propensity for normal-direction variations. At \(\langle H(N), \nu_\alpha \rangle = 0\), the submanifold corresponds to a minimal surface in the main manifold, characterizing an extremum of manifold-field coupling. The sign of \(V_{\text{eff}}\) determines the stability of this extremum, thus providing a stability criterion for the submanifold-matter system.

Specifically, \(V_{\text{eff}} =  - \frac{1}{2} \|\text{II}\|^2 \leq 0 \) implies the system occupies an unstable critical point, spontaneously evolving toward vanishing-curvature configurations to achieve geometric flatness—exemplified by Helicoid surfaces in \( \mathbb{R}^3\).

Remarkably, when \(\langle H(N), \nu_\alpha \rangle \neq 0\), stability persists if \(V_{\text{eff}} = 0\). This reveals a matter-submanifold interaction that maintains structural stability even for non-minimal submanifolds, as demonstrated by \( \mathbb{S}^2 \) (with radiu $R$) embedded in \( \mathbb{R}^3\).which satisfy \( \|H\|^2=\frac {4}{R^2}, \| \text{II} \|^{2}=\frac {2}{R^2} \)

Based on the preceding analysis, a direct consequence is that for deformable two-dimensional electronic systems with elliptical geometry embedded in Euclidean space, spontaneous evolution toward a spherical configuration is energetically favorable. 
We propose that the curvature of extra dimensions and the influence of different embedding methods on lower-dimensional electron dynamics may be observed through quantum simulation approaches analogous to those employed in Paper~\cite {PhysRevLett.129.120505,2022Quantum}. 
\subsection{Quantum mass dependent on additional dimensions}
Let us consider a more general manifold
M with \( \text{Ric}_M(\nu, \nu) \neq 0 \), and \(V_{\text{eff}} \neq 0 \). It can be shown that when the mean curvature vector is zero and the normal projection of the Ricci curvature of M is ‌positive‌, the particle-submanifold coupling admits ‌no stable critical points‌, cause of that Eq. \eqref{eq:secondoff} become:
\begin{equation} \label{eq:secondderivef}
\frac {\partial^2 f}{\partial {y}^2}= -\frac{1}{2}( \|\text{II}\|^2 + \sum_{1}^{m-n}\text{Ric}_M(\nu, \nu) )^2 \mathcal{O} \leq 0.
\end{equation}
This relation demonstrates that stable configurations of the system exist only when the Ricci curvature of the ambient manifold equals the negative squared norm of the second fundamental form of the embedded submanifold. Crucially, both quantities are intrinsically determined by the structure of the ambient manifold and the embedding characteristics of the submanifold. Such a constraint effectively introduces novel modes of stability. An example is hyperbolic plane \( {\Sigma}^2 \) embedded in \( \mathbb{H}^3 \) which led:
\[
\mathrm{Ric} = -2g=-2 , \| \text{II} \|^{2}=2.
\]
We conclude by examining the case where both the mean curvature \(\langle H(N), \nu_\alpha \rangle\) and the second fundamental form modulus\( \| \text{II}^\alpha \|^{2}\) vanish, indicating a ‌totally geodesic submanifold‌. For such configurations to exhibit stable criticality, the ambient manifold must be ‌Ricci-flat‌. A well-established exemplar of this principle is the ‌Calabi-Yau manifold~\cite{yau1978ricci},within whose framework we have demonstrated-to a certain extent-the stability of embedded ‌submanifold-matter field systems‌. This analysis illuminates the profound implications of ‌manifold-matter interactions‌ and provides critical insights into how ‌higher-dimensional geometries govern lower-dimensional physical phenomena‌. Subsequent investigations will further elucidate the emergent physical consequences through concrete applications.
Following the discussion of stability conditions, we now examine a specific case study to illuminate the profound physical implications embedded in our methodology-particularly regarding the substantive physical influence exerted by higher-dimensional manifolds on their submanifolds.
Without loss of generality and neglecting gravitational fields, 

{we set \( \mathbb{N} \) to be \( \mathbb{R}^3 \)—a model of the physical universe—and impose \( \mathbb{M}=\mathbb{R}^3 \times \mathbb{CH}^k \), which constitutes a canonical embedding wherein is a totally geodesic submanifold of \( \mathbb{R}^3 \times \mathbb{CH}^k \ \), satisfying \(\|\text{II}\|^2=0\).} which \( \mathbb{CH}^k \ \)is k dimensional Complex Hyperbolic Space, We substitute these specified conditions into the Klein-Gordon equation\eqref{eq:KG}.
\begin{equation}\label{eq:KGinCPn}
\begin{split}
\text{Ric}_{\mathbb CH^k} (\nu, \nu)= - \frac {2(k+1)}{r^2}. \\
K_{\text{eff}} = \frac{k(k+1)}{r^2}.
\end{split}
\end{equation}
where r is radius of curvature of complex Hyperbolic Space.
Let
\begin{equation}
m_g =\sqrt {K_{\text{eff}}}= \frac {\sqrt {k(k+1)}}{r}.
\end{equation}
Our theoretical framework demonstrates that \(k=0\) (corresponding to a massless scalar field) dynamically generates mass in excited states, scaling inversely with the curvature radius of uncompactified extra dimensions. Critically, the absence of compactification constraints admits arbitrary mass scales. This establishes a curvature-driven mass-generation mechanism distinct from Higgs-based approaches in degenerate geometries. 
Based on our derivation, we obtain a geometry-induced scalar field mass that inversely scales with the radius of curvature and exhibits dimension-dependent behavior. This outcome parallels the core mechanism of Kaluza-Klein reduction~\cite{kaluza1921unitatsproblem,klein1926quantentheorie}. Crucially, however, our framework operates ‌without requiring‌ additional periodic coordinate conditions ‌nor‌ defining scalar fields on extra dimensions. 
\subsection{‌Geometry-Induced Symmetry Breaking}
A more intriguing scenario emerges when considering the specific case of \(\mathbb{M}=\mathbb{R}^3 \times \mathbb{CP}^1 \), with the ricci curvature represent:
\begin{equation}\label{eq:cp1}
\text{Ric}_{\mathbb CP^1} (\nu, \nu)= \frac {4}{r^2},
\end{equation}
where incorporating this result into ‌the Lagrangian of a massless scalar field‌ yields non-trivial physical consequences.The Lagrangian density for a massless scalar field with $\phi^4$ interaction is given by:
\begin{equation}\label{eq:phi4}
\mathcal{L_M} = \frac{1}{2} (\partial_a \phi)^2 - {\lambda} \phi^4 ,
\end{equation}
Combining Eq. \eqref{eq:cp1} with analogous derivations, Eq. \eqref{eq:phi4} assumes the form:
massless scalar field with $\phi^4$ interaction is given by:
\begin{equation}\label{eq:phi4}
\begin{split}
&\mathcal{L_N} = \frac{1}{2} (\partial_i \phi)^2+{\mu_0}^2 \phi^2 - {\lambda} \phi^4 . \\
&{\mu_0}^2=\frac {2}{r^2}; V_{higgs}=-\frac {2}{r^2}\phi^2+ {\lambda} \phi^4.
\end{split}
\end{equation}
We show that geometric induction provides a mathematically natural pathway to spontaneous symmetry breaking, where the extrinsic curvature of spacetime intrinsically generates the required negative mass-squared term—eliminating the need for ad hoc potentials. This geometric origin not only explains the physical necessity of the higgs potential  but also makes the mechanism more intuitive: the negative term emerges naturally from spacetime deformation rather than being artificially imposed~\cite{higgs1966spontaneous}.

\subsection{‌Higgs Vacuum Near Low-Mass Black Hole Horizons}
Examining the Higgs field near the event horizon of a low-mass black hole modeled as an embedded surface in four-dimensional Minkowski spacetime, the Higgs potential in the small-mass limit takes the functional form derived in our approach. This result incorporates the geometric embedding of the Schwarzschild event horizon and its spacelike hypersurfaces (defined by constant Schwarzschild time \text{t}) within the ambient spacetime. The complete derivation is given in Appendix A. we derive the explicit expression for the Higgs potential‌: 
\begin{equation}\label{eq:horizon}
 V_{\phi}=-\frac {2}{r^2}\phi^2 - \frac{1}{r_s^2}\phi^2 + {\lambda} \phi^4 .  
\end{equation}
where $ r_s = 2GM $ is the Schwarzschild radius. 
The physical significance of this result lies in the fact that when the black hole mass \( M < 10^{11}  \mathrm{GeV} \), it induces a significant correction (greater than 1\%) to the Higgs mass. Consequently, a signature exceeding the standard Higgs mass becomes observable in the vicinity of the horizon of such low-mass black holes. 

Furthermore, by incorporating the analysis of metastable vacuum decay, the running coupling constant is given by the following expression~\cite{PhysRevD.99.024046}: 
\begin{equation}\label{eq:prd19}
\lambda_{\text{eff}} = g(\Lambda_{\phi}) \left\{ 
\left( \ln \frac{\phi}{M_p} \right)^4 - 
\left( \ln \frac{\Lambda_{\phi}}{M_p} \right)^4 
\right\}
\end{equation}

where $\lambda_{\text{eff}}$ represents the effective coupling constant, 
$g(\Lambda_{\phi})$ is a function of $\Lambda_{\phi}$,
$\phi$ and $\Lambda_{\phi}$ are physical quantities,
and $M_p$ denotes the Planck mass. 

When the black hole mass approaches the Planck mass, the corrected mass in Eq. \eqref{eq:horizon} \( \frac{1}{r_s}\propto M_{\mathrm{P}} \). This correction stabilizes the vacuum at the horizon, and Eq. \eqref{eq:prd19} confirms that \(\lambda_{\text{eff}} > 0 \). Consequently, Planck-scale micro black holes cannot trigger Higgs field decay to form decay bubbles.

\section{conclusion}
In our model, extra dimensions fundamentally ‌manifest as latent dynamical degrees of freedom‌ for the scalar field. The influence of higher-dimensional spacetime becomes encoded in the interactions between the submanifold (i.e., the 4D spacetime or 3D spacelike hypersurface) and the scalar field. We contend these results reveal ‌physically significant insights‌.

We emphasize the conceptual foundation of our model: A ‌freely embedded submanifold‌ within a higher-dimensional manifold induces ‌constraints on field dynamics‌ confined to the submanifold. This emerges intrinsically through interactions along the ‌normal directions‌ — not necessarily via an explicit normal-direction potential.

Employing differential geometry, we derive the equations of motion governing scalar fields confined to an n-dimensional embedded submanifold within a higher-dimensional ambient manifold. This intrinsically coupled manifold-field framework exhibits stability against perturbations in the extrinsic curvature, revealing how the ambient geometry governs submanifold physics. Crucially, without imposing compactification conditions, the analysis demonstrates the geometric generation of a scalar field mass term arising from the ambient curvature. 

Our model naturally incorporates a mechanism of spontaneous symmetry breaking via a geometrically induced Higgs potential, achieved without invoking compactification schemes. This demonstrates the precise physical implications of extra-dimensional models with inherent elegance.

We further establish Higgs vacuum stability in the near-horizon spacelike limit of microscopic black holes, proving the absence of vacuum decay near event horizons. Our calculations predict Higgs mass corrections in the vicinity of small-mass black holes—particularly those in terminal evaporation phases. 

These results demonstrate the theoretical robustness and broad applicability of our framework. A subsequent publication will address the interplay between non-compact extra dimensions and gauge symmetries.
Our framework maintains full compatibility with standard spacetime decompositions including the ADM formalism, requiring only metric signature adjustments for application to various relativistic scenarios. Future investigations will focus on three key directions: first, examining how higher-dimensional symmetries manifest in low-dimensional effective theories through dimensional reduction; second, exploring the rich geometric structures arising from submanifold-ambient manifold interactions; and third, extending the analysis to vector and spinor fields where new emergent phenomena beyond scalar interactions are anticipated.


\section*{Acknowledgments}
This work is jointly supported by the National Nature Science Foundation of China (Grants No. 11934008). R. C was funded by Army Engineering University of PLA(No. KYJBJKQTZQ23006).

\appendix
\section*{appendix A:schwarzschild horizon}

\renewcommand{\theequation}{\Alph{section}.\arabic{equation}} 
\setcounter{equation}{0}
\renewcommand{\theequation}{A\arabic{equation}}

The Schwarzschild metric in advanced null coordinates:
\begin{equation}
ds^2 = -\left(1 - \frac{r_s}{r}\right) dv^2 + 2 dv dr + r^2 (d\theta^2 + \sin^2\theta \, d\phi^2),
\end{equation}
where $ r_s = 2GM $ is the Schwarzschild radius. The event horizon is at $ r = r_s $.

The horizon $ \mathcal{H} $ is a null hypersurface at $ r = r_s $ with induced metric:
\begin{equation}
h_{AB} = r_s^2 \begin{pmatrix}
1 & 0 \\
0 & \sin^2\theta
\end{pmatrix}, \quad (A,B = \theta, \phi).
\end{equation}

The normal vector $ n^\mu = -\nabla^\mu v $ is null on the horizon. The extrinsic curvature is:
\begin{equation}
K_{AB} = \frac{1}{2} \mathcal{L}_n h_{AB} = \frac{1}{2} \partial_r h_{AB} \big|_{r=r_s}.
\end{equation}
Substituting $ h_{AB} = r^2 \mathring{h}_{AB} $ (unit sphere metric):
\begin{equation}
K_{AB} = r_s \mathring{h}_{AB} \implies K_{\theta\theta} = r_s, \quad K_{\phi\phi} = r_s \sin^2\theta.
\end{equation}

The norm is given by the trace:
\begin{equation}
\|K\| = \|\text{II}\| = h^{AB} K_{AB} = \frac{2}{r_s},
\end{equation}
using $ h^{AB} = r_s^{-2} \mathring{h}^{AB} $.


\normalem
\bibliographystyle{apsrev4-1}
\bibliography{references1}

\end{document}